\documentclass[12pt]{amsart}
\usepackage[english]{babel}
\usepackage{amsmath}
\usepackage{amssymb}
\usepackage{amsthm}
\usepackage{yhmath}
\usepackage{url}
\usepackage{graphicx}
\usepackage{caption}
\usepackage{enumitem}
\usepackage{algorithmicx,algpseudocode}
\usepackage[section]{algorithm}
\usepackage{caption}
\usepackage{wrapfig}
\usepackage{subcaption}
\usepackage{varwidth}
\newtheorem{theorem}{Theorem}[section]

\newtheorem{conjecture}[theorem]{Conjecture}

\theoremstyle{remark}

\usepackage{csquotes}

\DeclareMathSymbol{\minus} {\mathord}{operators}{"2D}

\usepackage[colorlinks=true, allcolors=blue]{hyperref}

\title{Fear,  Universality and Doubt in asset price movements}
\author{Igor Rivin}
\address{Mathematics Department, Temple University and The Cryptos Fund}
\email{rivin@temple.edu}
\date{\today}
\thanks{The author would like to thank Andrew P.~ Mullhaupt for interesting comments. All the computations in this paper were conducting using the \emph{Wolfram Mathematica} system. The stock price data came from \emph{Mathematica} itself (via the \textbf{FinancialData[]} mechanism, while the Bitcoin price history came from Yahoo! Finance. All data terminates on March 16 2018}
\begin{document}
\begin{abstract}
We take a look the changes of different asset prices over variable periods, using both traditional and spectral methods, and discover universality phenomena which hold (in some cases) across asset classes.
\end{abstract}
\maketitle

\section{Introduction}

There has been a considerable amount of effort dedicated to understanding (and, ultimately, predicting) asset prices. We will give an idiosyncratic overview in Section \ref{history}. 

In this note I describe a few experiments which both illustrate some of the shortcomings of the current dogma and show surprising universality phenomena. I do not pretend to understand why this universality holds.

Here is a brief outline of the rest of the paper.

In Section \ref{history} I give an overview of the current dogma. 

In Section \ref{primitive} I look at the distribution of returns of a number of secturities over varying lengths of time.

In Section \ref{spectral} I look at a somewhat more sophisticated way of computing statistics \emph{via} the Hankel matrix (\emph{trajectory matrix}) associated to the process. 
\subsection{Data} In this paper we use the following data sets (all sets end on the ides of March of 2018:
\begin{enumerate}
\item{Alphabet, Inc - stock ticker \textbf{GOOG}} We use data from the inception of the current version of the company in March of 2014 to the present.
\begin{enumerate}
%[label=aapl*)]
        \item Alphabet Inc Stock price, log scale\par
\begin{minipage}{\linewidth}
            \centering
            \includegraphics[width=10cm]{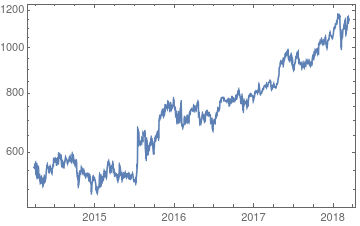}
            \captionof{figure}{Alphabet Inc Stock Price}
        \end{minipage}
        \end{enumerate}
\item{Apple Computer, Inc - stock ticker \textbf{AAPL}} We use data from 1991 to the present.
\begin{enumerate}
%[label=aapl*)]
        \item Apple Computer Stock price, log scale\par
\begin{minipage}{\linewidth}
            \centering
            \includegraphics[width=10cm]{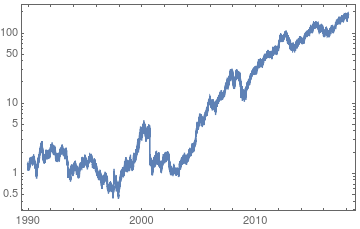}
            \captionof{figure}{Apple Stock Price}
        \end{minipage}
        \end{enumerate}
\item{General Electric, Inc -- stock ticker \textbf{GE}} We use data from 1962 to the present.
\begin{enumerate}
%[label=aapl*)]
        \item General Electric Stock price, log scale\par
\begin{minipage}{\linewidth}
            \centering
            \includegraphics[width=10cm]{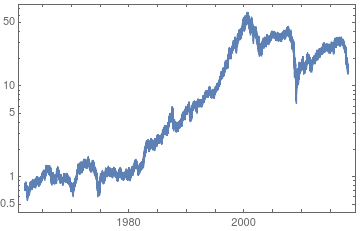}
            \captionof{figure}{GE Stock Price}
        \end{minipage}
        \end{enumerate}
\item{Bitcoin - ticker \textbf{BTC}} We use data from the time Bitcoin began trading in 2010 to the present day.
\begin{enumerate}
%[label=aapl*)]
        \item Bitcoin exchange rate, log scale\par
\begin{minipage}{\linewidth}
            \centering
            \includegraphics[width=10cm]{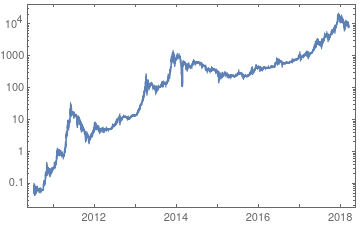}
            \captionof{figure}{Bitcoin exchange rate}
        \end{minipage}
        \end{enumerate}
\end{enumerate}
The three companies studied are all very large - this was intentional, since that meant that we would not have any periods of low liquidity. It should, however, be noted that before the decimalization in the early 2000s, the prices were discretized in much larger increments than later.
\section{A little history}
\label{history}
The first model of stock returns is due to Louis Bachelier \cite{bachelier1900theorie,bachelier1912calcul} has the following components. The first is that the returns on different days are independent and identically distributed (i.i.d.) and the second is that the distribution of returns is Gaussian. The first assumption stems from an early form of the Efficient Market Hypothesis (of which more anon), and the second from the philosophy that the price movement was due to a large number of small factors, and so some version of the Central Limit Theorem would make the move Gaussian. It should be noted that Bachelier anticipated the mathematical underpinnings for the study of Brownian motion (for which credit was later given to A.~Einstein).

Bachelier's work was forgotten in Finance (though not in Mathematics) circles, until it was rediscovered through the efforts of the University of Chicago group (Jimmie Savage in particular) in the 1950s and came back into vogue in the 1960. The model was modified to make not the returns themselves but their logarithms normally distributed. The usual explanation for this is that normally distributed returns might lead to asset prices themselves becoming negative -an impossibility. However, more sophisticated explanations are available, largely from the Kelly-ist contingent (of which the author considers himself a member), of which the founding member was D.~Bernoulli, who predated Kelly by a few hundred years - see the excellent compendium \cite{maclean2011kelly}. The efficient market hypothesis espoused by Bachelier was revived by Eugene Fama, Harry Markowitz, and Bill Sharpe (see, for example \cite{markowitz1952portfolio,fama1970efficient,sharpe1964capital}), all of whom later received Nobel Memorial Prizes in economics, while the alleged log normality of stock returns was used by Black and Scholes to derive their famous options pricing formula \cite{black1973pricing}. This formula was derived earlier (but not published) by Ed Thorp in the context of his work on warrants pricing \cite{thorp1967beat}, and three quarters of a century earlier by Bachelier (his hypotheses were slightly different, as pointed out above, but the predictive power of his model is not so different from Black-Scholes). Black and Scholes also received the Nobel memorial prize in economics for his efforts.\footnote{If the reader is keeping score, we are up to five Nobel prizes for Bachelier -- one in physics and four in Economics.}

Now, there are a couple of major problems with all of the above. 

First, it is quite clear that whatever the distribution of returns is, it is not log normal - if it were, market meltdowns like the ones in 1929, 1987, 2001, and 2008 would occur far less frequently than they actually do (since these events are several standard deviations away from the mean). This is usually explained away by saying that the distribution of returns is sort of log-normal, but with fatter tails. 

Secondly, the day to day returns are clearly \emph{not} independent. Table \ref{corrtab} shows autocorrelations of log returns of our four time series and also of the \emph{magnitudes} of the logs of autocorrelations.
\begin{table}
\caption{\label{corrtab}Correlations of returns and magnitudes of returns}
\begin{center}
    \begin{tabular}{| c |c | c |}
    \hline\hline
    Series & Log return autocorrelation & Absolute value of log return autocorrelation\\ \hline\hline
    GOOG & $0.0689368$ & $0.155424$ \\ \hline
    AAPL & $-0.00390584$ & $0.193075$ \\ \hline
   GE & $0.0165434$ & $0.2624213$  \\
    \hline
    BTC & $0.0292686$ & $0.384995$ \\ \hline
    \end{tabular}
 \end{center}
 \end{table}
 
 A look at Table \ref{corrtab} will tell us that while there might be some debate about the predictive power of the return on day $N$ of the return on day $N+1$ (although autocorrelation of $6\%$ as in the case of GOOG is certainly not negligible), there is absolutely \emph{no} doubt that the magnitude of the swing has some auto-regressive aspects. To deal with this problem, financial econometricians introduced another Brownian motion to control the volatility, and this work (more specifically, the definition of a GARCH stochastic process) has produced yet another Nobel Memorial Prize (this one to Robert Engle and Clive Granger). Interestingly, to the author's knowledge, the resulting technology is \emph{not} used by finance practitioners, due to its poor predictive power.
 
 \subsection{State of the art, as she is spoke}
 In the end, we have the following takeaways:
 \begin{enumerate}[label=\alph*)]
 \item The logarithms of the returns are modeled as Gaussian, except with fat tails.
 \item The returns on different days are kind of sort of independent random variables, but
 \item The return time series is not stationary ("heteroscedatic", because that sounds really cool).
 \end{enumerate}
 
 Not a very satisfying state of affairs.
 
\section{The distributions}
\label{primitive}
For our first attempt at enlightenment, we will do the simplest thing possible and look at the distributions of logs of daily returns of our chosen instruments. For each of them we will look at the histogram of returns and also at a kernel-smoothed version of the distribution (since it is conceivable that the discretization artifacts of the histogram will blind us to the truth.

\begin{enumerate}[label=\Alph*)]
        \item GE log return distribution\par
\begin{minipage}{\linewidth}
            \centering
            \includegraphics[width=15cm]{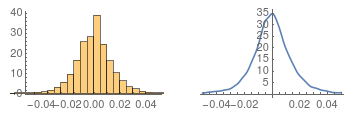}
            \captionof{figure}{GE log returns}
        \end{minipage}
        \item AAPL log return distribution\par
\begin{minipage}{\linewidth}
            \centering
            \includegraphics[width=15cm]{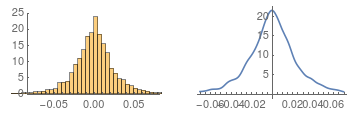}
            \captionof{figure}{AAPL log returns}
        \end{minipage}
        \item GOOG log return distribution\par
\begin{minipage}{\linewidth}
            \centering
            \includegraphics[width=15cm]{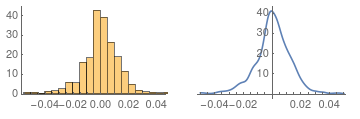}
            \captionof{figure}{GOOG log returns}
        \end{minipage}
        \item BTC log return distribution\par
\begin{minipage}{\linewidth}
            \centering
            \includegraphics[width=15cm]{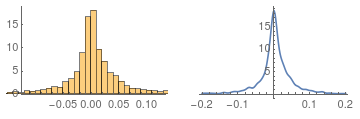}
            \captionof{figure}{Bitcoin log returns}
        \end{minipage}
        \end{enumerate}
        
A quick look at the graphs is enough to convince us that the only thing the distributions have in common with the Gaussian distribution is the unimodality and (rough) symmetry - in fact, the GOOG,AAPL, and BTC distributions are visibly \emph{not} symmetric about the mean. The one for GE is a bit more so.All of them have a more "triangular" shape around the mean than the Gaussian (which is relatively flat in the neighborhood of the origin). To check that the distributions are, indeed, very 
different from the Gaussian, we will look at the QQ(quantile-quantile) plots. For two continuous distributions $d_1, d_2$ the points in such a plot are the points $(c_1(q), c_2(q)),$ where $c_1(q)$ is the inverse of the cumulative distribution function of $d_1$ applied to $q$ and $c_2$ is the inverse of the cumulative distribution function of $d_2$ likewise applied to $q.$ If $d_1 = d_2,$ the qq plot will simply be the line $x=y.$ If $d_1$ and $d_2$ differ by scale and or location (that is, $d_1(x) = d_2(a x + b)$) the qq plot will be a straight line. If the right tail of $d_1$ is fatter than that of $d_2,$ the graph will veer higher than the $x=y$ line, and similarly for the left tail.

\begin{figure}
\centering
\includegraphics[width=0.6\textwidth]{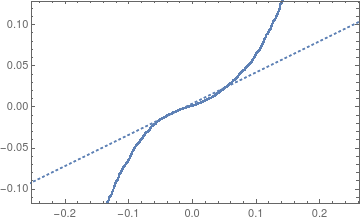}
\caption{\label{btcquant} Q-Q plot of Bitcoin log returns vs the standard Gaussian.}
\end{figure}

\begin{figure}
\centering
\includegraphics[width=1.0\textwidth]{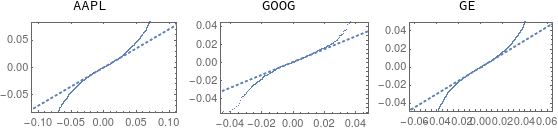}
\caption{\label{gaussgrid} Q-Q plot of AAPL, GOOG, and GE log returns vs the standard Gaussian.}
\end{figure}

As a warm-up we will look at the qq plot of Bitcoin log-returns vs the standard Gaussian $\mathcal{N}(0, 1)$ - see Figure \ref{btcquant}.
It is quite clear that our original impression was correct, and the Bitcoin returns are nothing like log-normal. However, Bitcoin is a new instrument, so who can tell what is going on with it, so let's look at our equity stalwarts - see Figure \ref{gaussgrid}.
We see again that these returns are nothing like log-normal.

\begin{figure}
\centering
\includegraphics[width=1.0\textwidth]{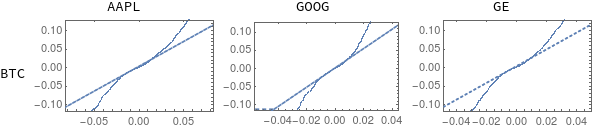}
\caption{\label{btcgrid} Q-Q plot of AAPL, GOOG, and GE log returns vs Bitcoin log returns.}
\end{figure}

Let us see if the Bitcoin returns really are qualitatively different from the large cap stock returns. We see the Q-Q plots in Figure 
\ref{btcgrid}.
We see that our conjecture about the idiosyncrasies of Bitcoin returns was justified. Finally, let us look at how differently distributed the returns of our large cap triad are. The results are in Figure \ref{qgrid2}:

\begin{figure}
\centering
\includegraphics[width=1.0\textwidth]{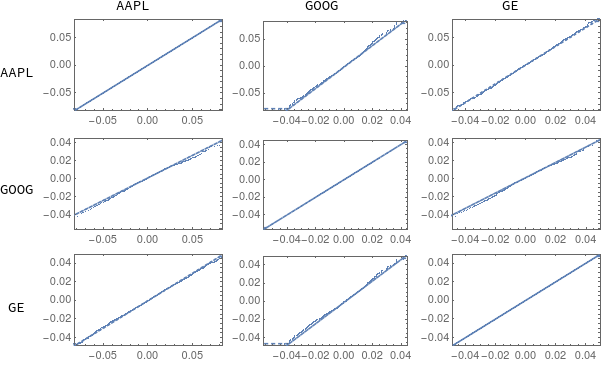}
\caption{\label{qgrid2} Q-Q plots of AAPL, GOOG, and GE log returns vs each other.}
\end{figure}
Now we are in for a bit of a surprise: the distributions are essentially the same, up to scaling and location. This is our first universality:
\begin{conjecture}[First Universality Conjecture]
The distributions of (at least large cap) equities are essentially the same, and differ only in scale and location.
\end{conjecture}
The study of daily returns distribution is an inherently \emph{static} activity, and so we must look for more sophisticated tools to look at the dynamics of the situation.

\section{The trajectory matrix and its spectrum}
\label{spectral}
Suppose we want to predict stock prices (who doesn't?) and we want to apply some out of the box machine learning algorithm. If all we have are past stock prices, then the usual way to approach it is to structure our data as follows: Let us assume that our series of log returns is $\mathcal{X} = X_0, X_1, \dotsc, X_n \dotsc.$ If our look back horizon is $k$ time periods, then our inputs are laid out as the matrix:
\[H_{k, n} = \begin{pmatrix}
X_1 & X_2 &\dotsc & X_k\\
X_2 & X_3  &\dotsc & X_{k+1}\\
\adots & \adots & \dotsc & \adots\\
X_n & X_{n+1} & \dotsc & X_{k+n-1}
\end{pmatrix},
\]
while our outputs are just the vector $(X_{k+1}, \dotsc, X_{n+k}.$
The matrix $H_{k,n}$ is known in the time series community as the \emph{trajectory matrix}. It is a Hankel matrix (meaning that the $ij$-th element of the matrix depends only on $i+j$), and it is reasonably clear that its singular values will have some relationship with the properties of the time series $\mathcal{X}.$ Indeed, the study of this relationship is a whole area of time series analysis - this subarea is known as SSA - Singular Spectrum Analysis. See, for example \cite{golyandina2013singular}. Singular spectrum analysis generally deals with $k \ll n,$ and is used for divining \emph{local} properties of the time series, but we will take a bird's eye view, and make $k$ very large. So large that the matrix will be square, with both $n$ and $k$ being half as large as our entire history. We will call this matrix $\mathcal{H}.$ We will study the spectrum of $\mathcal{H}$ and see if we find anything interesting. 
\subsection{Eigenvalues of Random Matrices} Now, it should be noted that random matrix theory is a huge subject, and there are generally two kinds of invariants to look at: bulk invariants (the spectral measure of our favorite class of asymptotically very large matrices), and fine invariants (these include the spacings between adjacent eigenvalues, or the behavior of eigenvalues on the "edge" of the spectrum. It is often true that fine invariants are more universal than bulk invariants (for example, there are strong connections between spacings between eigenvalues of very large Hermitian matrices and spacings between zeros of the Riemann zeta function - see, for example the classic book by N. Katz and P. Sarnak - \cite{katz1999random}. In any case, nothing much seems to be known for spacings of large symmetric Hankel matrices, but there are results on the limiting spectral measure by Bryc, Dembo, and Jiang - \cite{bryc2006spectral} - the spectral measure of a large square Hankel matrix coming from a series of i.i.d. random variables (with finite variance) $\mathcal{X}$ does approach a limit, which is a funny looking bimodal distribution, though aside from its general shape, it seems that not so many properties of this distribution are known. Of course, we can easily sample from it, by generating a large sample from the standard Gaussian distribution, constructing the trajectory matrix, and computing its eigenvalues.\footnote{eigenvalues of Hankel matrices can actually be computed in time $O(n^2 \log n).)$} Here are the results:
\begin{enumerate}[label=\Alph*)]
        \item i.i.d. Gaussian trajectory matrix ($3000\times 3000$) eigenvalue bulk distribution\par
\begin{minipage}{\linewidth}
            \centering
            \includegraphics[width=\textwidth]{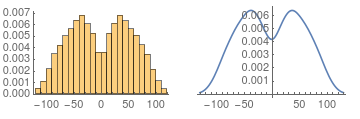}
            \captionof{figure}{Random Hankel Matrix spectral density}
        \end{minipage}
        \item i.i.d. Gaussian trajectory matrix ($3000\times 3000$) eigenvalue spacing distribution.
        \begin{minipage}{\linewidth}
            \centering
            \includegraphics[width=\textwidth]{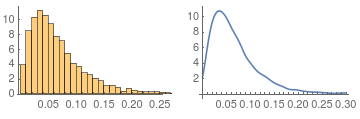}
            \captionof{figure}{Random Hankel Matrix spacings distribution}
        \end{minipage}
\end{enumerate}
The careful reader will see that the spacings distribution is showing the characteristic ``level repulsion'' phenomenon - the peak of the distribution is away from zero, just as it is for the popular GOE, GUE, and GSE ensembles of matrices -- \cite{mehta2004random}. However, the distribution is none of those three. Below are some diagrams which drive the point home:
\begin{enumerate}[label=\Alph*)]
        \item Spacing distributions. \par
\begin{minipage}{\linewidth}
            \centering
            \includegraphics[width=\textwidth]{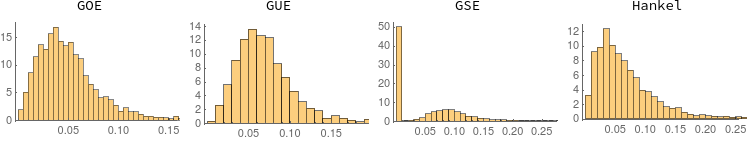}
            \captionof{figure}{Spacing distributions for standard ensembles and random Hankel Matrices}
        \end{minipage}
        \item i.i.d. Gaussian trajectory matrix ($3000\times 3000$) eigenvalue spacing distribution.
        \begin{minipage}{\linewidth}
            \centering
            \includegraphics[width=\textwidth]{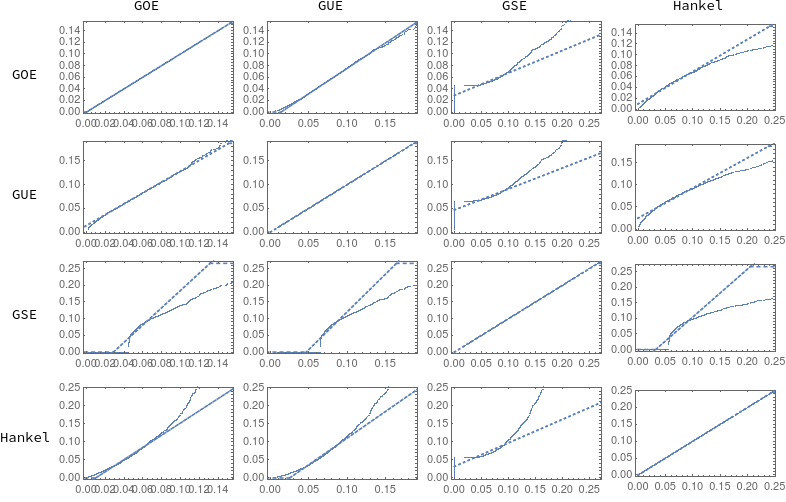}
            \captionof{figure}{pairwise quantile-quantile plots of spacing distributions}
        \end{minipage}
\end{enumerate}

The sharp-eyed observer will note that the Hankel spacing distribution has fatter tails than the other ones.
\subsection{Back to the stocks}
Our journey is now nearing the end. For our next act, we generate maximal trajectory matrices for our data sets, and see what happens.

\begin{figure}
\centering
\includegraphics[width=1.0\textwidth]{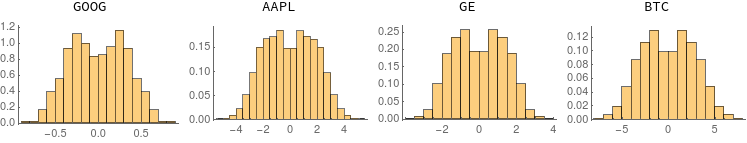}
\caption{\label{coeig} Bulk eigenvalue distribution of trajectory matrices.}
\end{figure}

First, the bulk eigenvalue distributions (Figure \ref{coeig})

\begin{figure}
\centering
\includegraphics[width=1.0\textwidth]{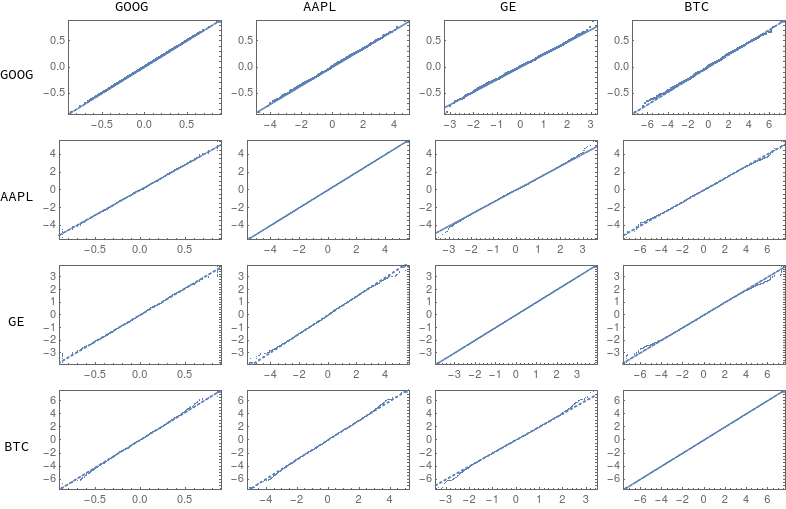}
\caption{\label{coqeig} Bulk eigenvalue distribution of trajectory matrices.}
\end{figure}

How close are these to the ``mother'' random Hankel distribution? A look at Figure \ref{coqeig} tells us that they are clearly different from their parent (though less different in the case of GE, which is a sign that the differences might be a function of not-quite large enough sample size.)

\begin{figure}
\centering
\includegraphics[width=1.0\textwidth]{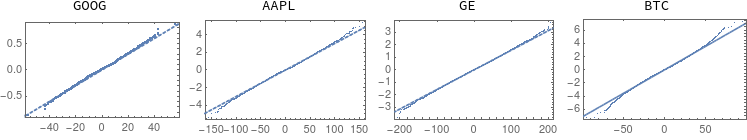}
\caption{\label{releig} Bulk eigenvalue distribution of trajectory matrices.}
\end{figure}
 The spacings are reasonably  those of the parent distribution, as shown below. Note that BTC is no longer an outlier.
\begin{enumerate}[label=\Alph*)]
\item Spacing distributions of trajectory matrices \par
\begin{minipage}{\linewidth}
            \centering
            \includegraphics[width=\textwidth]{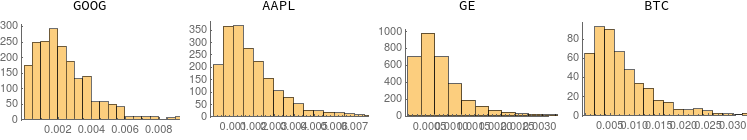}
            \captionof{figure}{Spacing distributions}
        \end{minipage}
\item Quantile plots of spacing distributions of trajectory matrices vs random Hankel\par
\begin{minipage}{\linewidth}
            \centering
            \includegraphics[width=\textwidth]{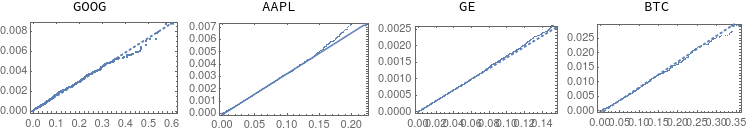}
            \captionof{figure}{Spacing distributions vs random Hankel}
        \end{minipage}
\end{enumerate}

Finally, we see how the spectra of the various trajectory matrices compare. First, the bulk distributions (Figure \ref{coqeig})

And then the spacings (Figure \ref{coqspacings}):
\begin{figure}
\centering
\includegraphics[width=1.0\textwidth]{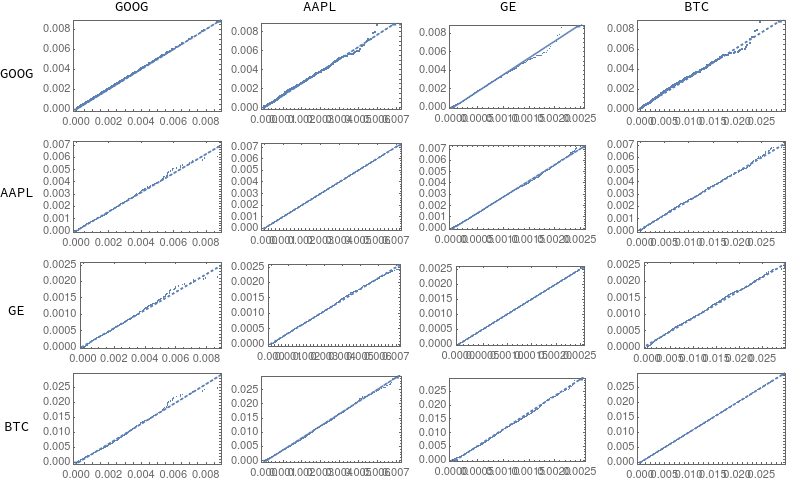}
\caption{\label{coqspacings} Eigenvalue spacing distribution of trajectory matrices.}
\end{figure}
The bulk distributions look very close, the spacings distributions a little less so, but we have again eliminated the apparently large differences between instruments (even going across asset classes). This leads us to another conjecture:
\begin{conjecture}[spectral universality]
The spectral distribution and the spectral spacing distribution of trajectory matrices exhibits universality, across asset classes.
\end{conjecture}
\bibliographystyle{alpha}
\bibliography{sample}

\begin{thebibliography}{MTZ11}

\bibitem[Bac00]{bachelier1900theorie}
Louis Bachelier.
\newblock {\em Th{\'e}orie de la sp{\'e}culation}.
\newblock Gauthier-Villars, 1900.

\bibitem[Bac12]{bachelier1912calcul}
L~Bachelier.
\newblock Calcul des probabilites, 1912.

\bibitem[BDJ06]{bryc2006spectral}
W{\l}odzimierz Bryc, Amir Dembo, and Tiefeng Jiang.
\newblock Spectral measure of large random hankel, markov and toeplitz
  matrices.
\newblock {\em The Annals of Probability}, pages 1--38, 2006.

\bibitem[BS73]{black1973pricing}
Fischer Black and Myron Scholes.
\newblock The pricing of options and corporate liabilities.
\newblock {\em Journal of political economy}, 81(3):637--654, 1973.

\bibitem[Fam70]{fama1970efficient}
Eugene~F Fama.
\newblock Efficient capital markets: A review of theory and empirical work.
\newblock {\em The journal of Finance}, 25(2):383--417, 1970.

\bibitem[GZ13]{golyandina2013singular}
Nina Golyandina and Anatoly Zhigljavsky.
\newblock {\em Singular Spectrum Analysis for time series}.
\newblock Springer Science \& Business Media, 2013.

\bibitem[KS99]{katz1999random}
Nicholas~M Katz and Peter Sarnak.
\newblock {\em Random matrices, Frobenius eigenvalues, and monodromy},
  volume~45.
\newblock American Mathematical Soc., 1999.

\bibitem[Mar52]{markowitz1952portfolio}
Harry Markowitz.
\newblock Portfolio selection.
\newblock {\em The journal of finance}, 7(1):77--91, 1952.

\bibitem[Meh04]{mehta2004random}
Madan~Lal Mehta.
\newblock {\em Random matrices}, volume 142.
\newblock Elsevier, 2004.

\bibitem[MTZ11]{maclean2011kelly}
Leonard~C MacLean, Edward~O Thorp, and William~T Ziemba.
\newblock {\em The Kelly capital growth investment criterion: Theory and
  practice}, volume~3.
\newblock world scientific, 2011.

\bibitem[Sha64]{sharpe1964capital}
William~F Sharpe.
\newblock Capital asset prices: A theory of market equilibrium under conditions
  of risk.
\newblock {\em The journal of finance}, 19(3):425--442, 1964.

\bibitem[TK67]{thorp1967beat}
Edward~O Thorp and Sheen~T Kassouf.
\newblock {\em Beat the market: a scientific stock market system}.
\newblock Random House, 1967.

\end{thebibliography}

\end{document}